\documentclass[runningheads]{svmult}

\usepackage{makeidx}   
\usepackage{graphicx}  
\usepackage{subeqnar}  
\usepackage{multicol}  
\usepackage{physprbb}  
\makeindex             

\begin{document}
\title*{Star Formation in High Redshift Radio Galaxies}
\toctitle{Star Formation in High Redshift Radio Galaxies}
%
%
\titlerunning{Star Formation in High Redshift Radio Galaxies}
%
\author{Carlos De Breuck\inst{1}
\and Michiel Reuland\inst{2}
}
\authorrunning{Carlos De Breuck et al.}
%
\institute{ESO, Karl Schwarzschild Stra\ss e 2, D-85748 Garching, Germany
\and Sterrewacht Leiden, Postbus 9513, 2300 RA Leiden, The Netherlands}

\maketitle              

\begin{abstract}
We present evidence for star formation in distant radio galaxies based on deep FORS1/VLT spectropolarimetry and SCUBA/JCMT 850$\mu$m photometry.
\end{abstract}

High redshift radio galaxies are efficient tracers of massive stellar systems. Their powerful radio emission requires massive black holes ($>10^9$M$_{\odot}$), which are known to reside in massive ellipticals [e.g. 5]. An attempt to compare the stellar mass of radio galaxies with other galaxies can be made in the Hubble $K-z$ diagram. This shows that radio-selected galaxies have brighter $K-$band magnitudes than $K-$band selected galaxies [2], which can be interpreted in terms of radio galaxies having baryonic masses as high as 10$^{12}$M$_{\odot}$, i.e. more than an order of magnitude higher than a typical HDF galaxy at $z\sim 3$ [7].

A serious complication is the presence of an AGN, which may also contribute to the {\it observed} $K-$band emission. A direct way to determine the contribution of scattered AGN emission is spectropolarimetry. [8] have found {\it rest-frame} UV polarizations $P>$10\% in five out of 10 $z\sim$2.5 radio galaxies, indicating scattered light from a burried AGN. However, at $z>3$, 2 out of 3 radio galaxies with spectropolarimetry information have $P<$5\% [4,8]. One of these (4C~41.17, $z=3.8$) shows stellar photospheric absorption lines [3], providing direct evidence that the UV continuum is dominated by light from young, hot stars.

To verify whether the contribution of young stars to the UV continuum rises with redshift, we have observed four $z>3$ radio galaxies with FORS1/VLT. All targets have been selected to have $22<R<23$ to ensure sufficient S/N to determine the polarization in integration times of 6 to 14 hours per target (De Breuck et al. in preparation). All four galaxies show $P\le$8\%, and two $P\le$4\%. In particular, we detect the CIII~$\lambda$~1428~\AA\ photosheric absorption line in one of the unpolarized sources, TN~J2007$-$1316 at $z=3.84$ (Fig. 1).

Another indication of star formation is submm dust emission. If the dust is heated by massive stars rather than directly by the AGN, this implies star formation rates as high as a few 1000M$_{\odot}$yr$^{-1}$. [1] and [6] have obtained SCUBA 850~$\mu$m photometry of 69 $z>1$ radio galaxies. The detection rate and 850$\mu$m luminosity increase dramatically at $z>3$, suggesting higher star formation rates. 

Figure~2 shows that the UV polarization and dust emission appear anti-correlated. This can be interpreted as evidence that these sources selected to have $R<23$, require either an important scattered quasar component (hence high polarization), or (if unpolarized), a high contribution from a young stellar population, which can heat the dust in these objects. A more detailed analysis will be presented by Reuland et al. (2004).

\begin{figure}[ht]
\begin{center}
\begin{tabular}{cc}
\includegraphics[width=.55\textwidth,height=.4\textwidth,angle=90]{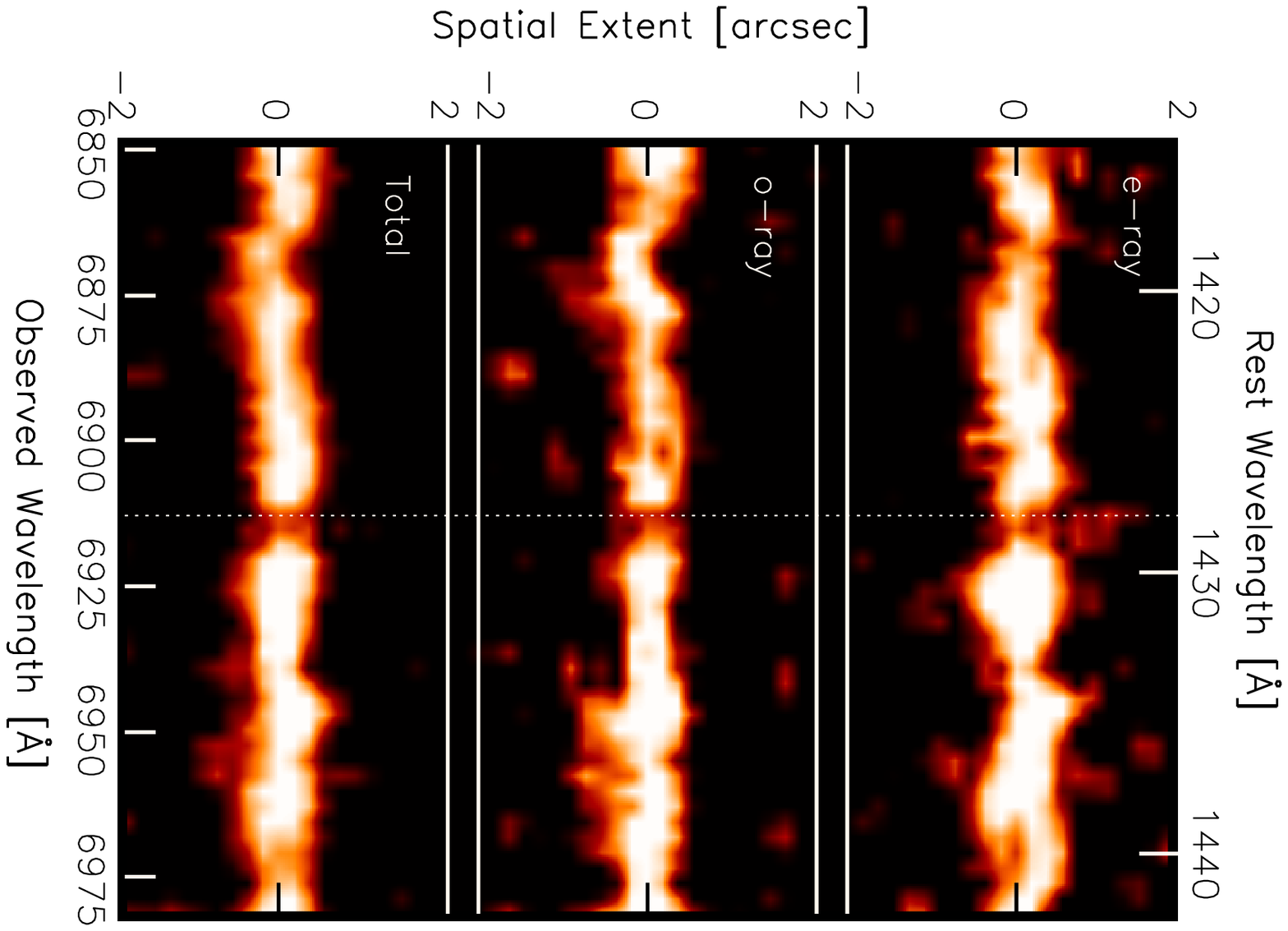}&
\includegraphics[width=.55\textwidth,height=.5\textwidth]{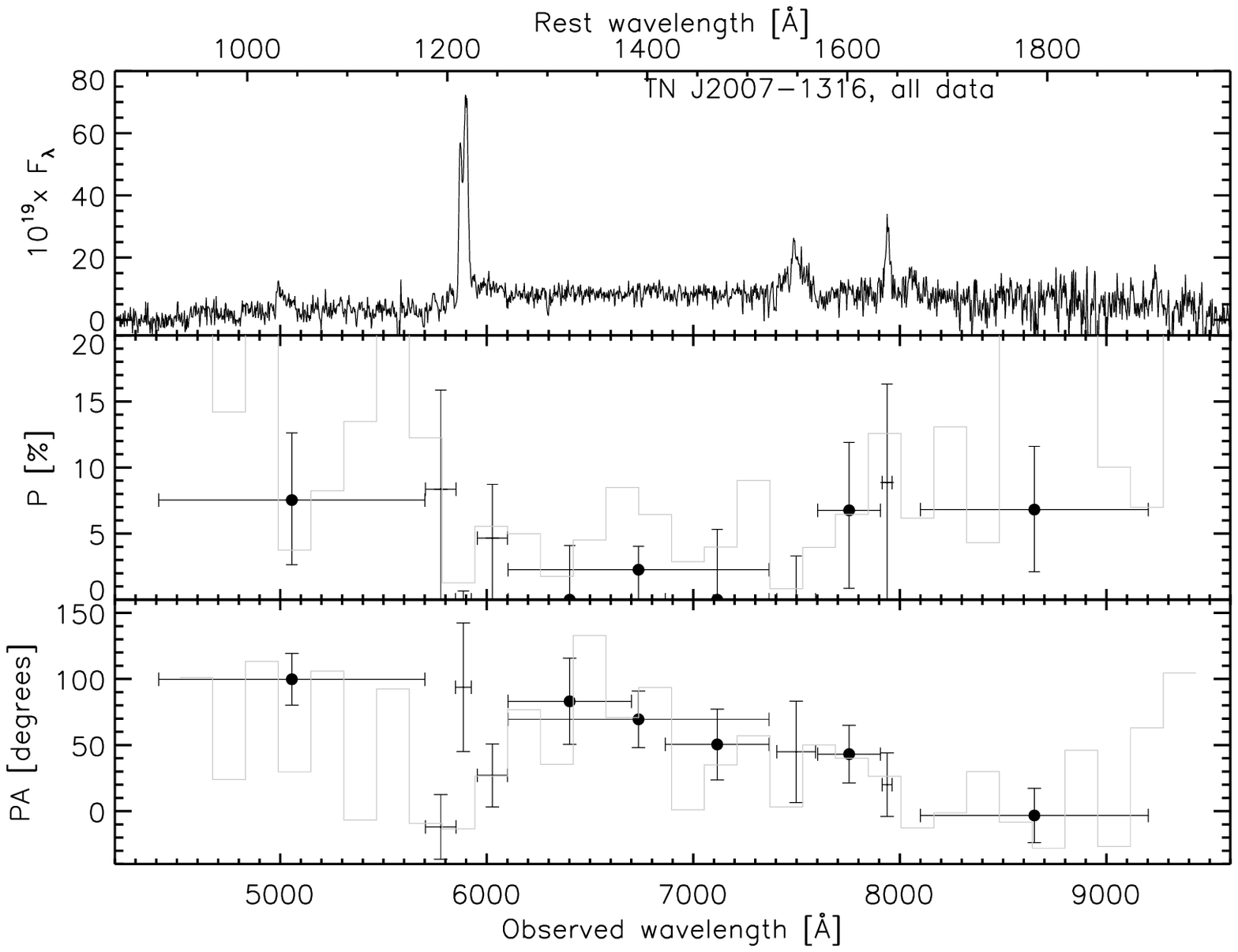}\\
\end{tabular}
\end{center}
\vspace{-0.3cm}
\caption[]{VLT/FORS1 spectropolarimetry of TN~J2007$-$1316:
{\it Left:} Part of the summed extraordinary ray, ordinary ray, and total intensity 2-dimensional spectra. The dotted line indicates the wavelength of the CIII~$\lambda$~1428~\AA\ photospheric absorption line, assuming the redshift of the AGN.
{\it Right:}
Total intensity, percentage polarization, and polarization angle. The grey histograms show the data with a 150~\AA\ binning.}
\label{eps1}
\end{figure}

\begin{figure}[ht]
\begin{center}
\includegraphics[width=.4\textwidth]{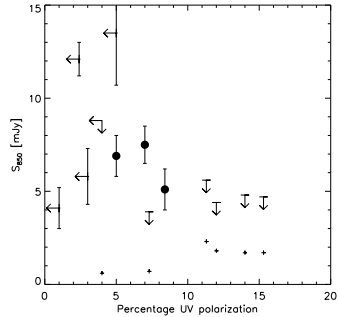}
\end{center}
\vspace{-0.3cm}
\caption[]{Anti-correlation between UV continuum polarization and $S_{850}$ for $z>1$ radio galaxies. For non-detections in $S_{850}$, we show nominal detections (+) and 3$\sigma$ limits.}
\end{figure}

%

\end{document}